# Symbolic Techniques for Deep Learning: Challenges and Opportunities

Belinda Fang[1, 4], Elaine Yang[2, 4], and Fei Xie[3]
[1]University of California Los Angeles, CA 90024, USA
bfang@g.ucla.edu
[2]University of Southern California, CA 90007, USA
etyang@usc.edu
[3]Portland State University, OR 97291, USA
xie@pdx.edu
[4]Paper written during summer internships at Portland State University

**Abstract**

As the number of deep learning frameworks increase and certain ones gain popularity, it spurs the discussion of what methodologies are employed by these frameworks and the reasoning behind them. The goal of this survey is to study how symbolic techniques are utilized in deep learning. To do this, we look at some of the most popular deep learning frameworks being used today, including TensorFlow, Keras, PyTorch, and MXNet. While these frameworks greatly differ from one another, many of them use symbolic techniques, whether it be symbolic execution, graphs, or programming. We focus this paper on symbolic techniques because they influence not only how neural networks are built but also the way in which they are executed.

To understand how symbolic techniques play a role in machine learning, we examine TensorFlow, Keras, and MXNet which all offer symbolic APIs that provide an interface for symbolic programs and execution. With symbolic APIs, models have a graph-like structure and execution treats program variables as symbols, allowing the program to reason about multiple concrete inputs in a single execution. In the case of deep neural networks, symbolic execution is able to obtain the typically hidden mathematical characterizations of internal behavior.

Out of the frameworks surveyed, PyTorch is the only one that is completely imperative, or non-symbolic. PyTorch exemplifies an imperative framework and is useful in comparing and contrasting the benefits of using and forgoing symbolic techniques. In studying imperative frameworks, we observe that it is often more user friendly and flexible, but lacks the optimization opportunities available to symbolic frameworks.

Limitations of symbolic techniques have led to efforts in integrating symbolic and nonsymbolic aspects in deep learning, opening up new possibilities for symbolic techniques. For example, the Gluon API by Apache MXNet bridges the gap between imperative programming and symbolic execution through hybridization. Frameworks such as JANUS attempt to translate imperative programs into symbolic graphs, while approaches like DeepCheck attempt to use symbolic execution to analyze and validate imperative neural network programs. Symbolic analysis has also been paired with concrete execution in a technique called concolic testing in order to better test deep neural networks. Our study of these developments exemplifies just a few of the many ways the symbolic techniques employed by popular frameworks have the opportunity to be altered and utilized to achieve better performance.

# 1 Introduction

Machine learning (ML) [1, 2, 3, 4, 5] is a rapidly evolving field of computer science and an increasingly evident part of our world today. Self driving cars, GPS traffic estimations, online fraud detection, and video surveillance are just some of the many ways in which the power of machine learning has been leveraged.

Machine learning, a subfield of artificial intelligence, involves the use of computer algorithms to enable systems to autonomously learn from data. The goal with machine learning is for computers to be able to learn and adjust their algorithms with little or no human intervention.

Deep learning (DL) [1, 2, 3] is a subset of machine learning that trains the model with hidden layers that operate between the input and output. Thus, while both machine and deep learning involve the creation of algorithms, deep learning utilizes multiple levels of algorithms that each interpret data differently. While artificial neural networks are made up of an input and output layer and at least one hidden layer in between, deep neural networks are characterized by having multiple hidden layers and are the focus of deep learning. For example, when tasked with identifying images of dogs and cats into their respective categories, traditional machine learning algorithms would require training images to be labeled by predefined features whereas deep learning would send the images through its layers of neural networks which would hierarchically determine image characteristics. Thus, deep neural networks allow machines to be able to train themselves to perform a task.

This paper surveys the deep learning frameworks, TensorFlow [11, 17, 18, 25, 26], Keras [21, 24, 25, 26], PyTorch [19, 20, 22, 23], and MXNet [27, 29] to explore how symbolic techniques are utilized in the context of deep learning models. The investigation considers symbolic APIs, programming, representation, analysis, and execution to understand the advantages and disadvantages of symbolic techniques in deep learning and the challenges and opportunities that can come from them.

Throughout this survey, we observe that many deep learning models only take on certain symbolic aspects, demonstrating how a symbolic approach is not a one-size-fits-all solution and revealing how frameworks attempt to utilize symbolic techniques without experiencing their limitations. While some frameworks are completely symbolic or imperative, there are attempts, such as the Gluon API [6, 7] by MXNet, that take aspects from both paradigms. Gluon stands out from the other three surveyed frameworks for its hybrid approach that allows users to develop and build models in an imperative fashion but execute symbolically, eliminating the difficulties of writing symbolic code but obtaining the performance optimizations from symbolic execution.

Finally, surveying various deep learning frameworks highlighted the challenges in dealing with symbolic techniques. Many imperative frameworks try to use symbolic techniques to obtain the

Page 2

optimizations and analysis provided by symbolic frameworks and execution. However, converting imperative programs into symbolic graphs is challenging. For example, if a user wants to use symbolic execution on an imperative Python program, certain Python characteristics, such as variable type, control flow decisions, and the values to read from or write to the heap, would be required to construct the symbolic graph but cannot be determined until runtime [8]. Furthermore, these characteristics are not guaranteed to remain the same after graphs are generated, therefore the resulting graphs cannot always be reused [8]. Attempts to combine the two paradigms often require that users provide the necessary information or generate erroneous results when converting an imperative program to symbolic graphs [8]. Another difficulty with symbolic techniques is that building a graph and executing it are two separate entities, and developing an efficient symbolic execution engine for neural networks is not an easy task. Neural networks [9] often have no branching and are non-linear. In addition, they often consist of thousands of neurons that make it difficult to scale down to fit the capabilities of the currently available symbolic reasoning tools [9].

## 2 Background

Symbolic and imperative programming are two paradigms that developers often choose between when building programs. It is important to understand the two styles in our survey of deep learning models, because many models choose to utilize one or both paradigms. Looking at the way in which they do so highlights where symbolic may be superior or inferior to imperative in the context of deep learning.

**Imperative programming.** Imperative programming [10, 43] is the oldest programming paradigm, meaning most programmers will learn an imperative language early on in their training. However, for those who are not familiar with the imperative style, it is known to be relatively easy to pick up. This ease of use and support of native language features contribute to this style's popularity.

Imperative APIs provide an interface for imperative programming:

```
1   import numpy as np
2   a = np.ones(10)
3   b = np.ones(10)
4   c = b * a
5   d = c + 1
```

Figure 1: Imperative program in NumPy

With the imperative program in Figure 1, executing the mathematical operations in line 4 and 5 would run the actual numerical computations and store the results in their respective variables. This is because operations are immediately executed in imperative programs, as there is no separate graph construction step before execution. With imperative programs, tasks are performed step by step through a guided sequence that changes the program state. The imperative paradigm is closely related to the actual machine and its architecture, making the programmer closer to the machine.

**Symbolic programming.** Certain deep learning models use symbolic APIs [10, 11] which view the model as a hierarchical, graph-like structure. With symbolic APIs, programs first compile, then become executable, and finally bind actual input values to get the computation result. Symbolic APIs provide an interface for symbolic programming and configuring symbolic graphs.

With symbolic programming [16], functions are defined in terms of placeholders. Translating the imperative style code from Figure 1 to a symbolic style program would look as follows:

```
1    A = Variable('A')
2    B = Variable('B')
3    C = B * A
4    D = C + Constant(1)
5    f = compile(D)
6    d = f(A=np.ones(10), B=np.ones(10)*2)
```

Figure 2: Symbolic style program

In Figure 2, when line 3 is reached, a symbolic graph is generated to represent the computation, but no actual computation takes place yet. Symbolic programs are characterized by this separation of building and executing the symbolic graph [15, 16].

When the program is executed, it goes through symbolic execution [12, 13, 14], a frequently used technique in the field of computer science that is known for its effective ability to detect infeasible paths, generate test inputs, find bugs and vulnerabilities, prove equivalent code segments, generate program invariants, and repair programs. Symbolic execution was first introduced in the 1970s, but wasn't widely used until developments in SAT/SMT solving and applications of Moore's Law made it more practical [13]. Now, symbolic execution is commonly used in program analysis, research, and industry tools. Some popular symbolic execution implementations include KLEE [45], Java PathFinder [46], SAGE [47], and Jalangi [48].

Symbolic execution abstractly executes a program by treating inputs as symbols rather than as concrete values. The program executes on symbolic values, maps variables to these values, creates path conditions that encode the branch decisions, and forms an execution tree using all the program's paths. By doing this, inputs are classified by their execution paths through the code and one execution can cover a number of possible inputs that share the same path. For example, if we wanted to check whether variable D from line 4 in Figure 2 is greater than zero, symbolic execution would view inputs as groups that would either satisfy or fail the condition.

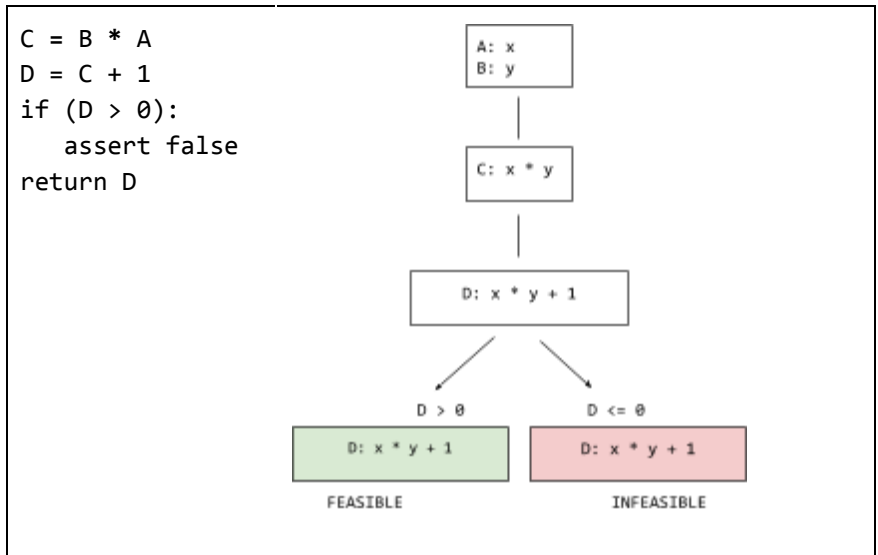

Figure 3: Code snippet from Figure 2 and its symbolic execution tree

In Figure 3, the program maps variables A, B, C, and D to symbolic values, x and y. When the program executes, path conditions are formed into an execution tree and paths are classified as feasible and infeasible. Because input values are represented by symbolic constants during execution, testing can be generalized and multiple test runs can be simulated for one path. Thus, symbolic execution is able to analyze a program by identifying how different inputs impact the execution of each part of a program.

**Symbolic graph execution.** Symbolic graph execution [15] is evident in deep learning frameworks such as TensorFlow, Caffe2, and MXNet and involves constructing neural networks as symbolic dataflow graphs. The graph's vertices correspond to the network's states and operations, and its edges show how data flows between vertices. By having a graph representation where operations are carried out as their dependents are solved, frameworks can find places to run operations in parallel and perform optimizations such as common subexpression elimination or constant folding.

For example, the following code in MXNet's Python Symbolic API configures a computation (symbolic) graph that is illustrated to the right of the code.

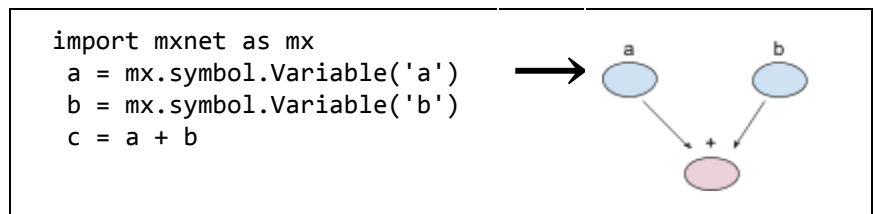

Figure 4: Symbolic graph representation of an MXNet program

# 3 Survey of Deep Learning Models

Symbolic DL frameworks such as TensorFlow and Keras build graphs to represent computation before execution, while imperative frameworks such as TensorFlow Eager and PyTorch directly execute programs without a separate optimization phase. MXNet Gluon attempts to combine the two styles to obtain the execution efficiency of symbolic frameworks but also the more intuitive programming approach apparent in imperative frameworks.

|  | Framework Description | Symbolic Techniques | Imperative Techniques |
|---|---|---|---|
| TensorFlow [11, 17, 18, 25, 26] | TensorFlow is an open-source symbolic math library for neural networks and dataflow programming developed by Google. TensorFlow upgraded to version 2.0 which adopted Keras as a high level API. | TensorFlow builds symbolic graphs to represent computation before execution. TensorFlow 1.x only supports static computation graphs which must be executed via sessions. Symbolic execution in TensorFlow means that computations are not performed until data is injected. | Users can enable eager (imperative) execution in TensorFlow 1.x, while TensorFlow 2.0 (TensorFlow Eager) is imperative by default. Imperative execution involves no graph construction and allows the framework to support dynamic models. |
| Keras [21, 24, 25, 26] | Keras is a high-level interface written in Python that runs on backend frameworks like TensorFlow. It is designed to be easy to use and allow for fast experimentation with deep neural networks. | Keras Sequential and Keras Functional provide symbolic APIs where models are translated into graphs with symbol representation. Keras uses symbolic tensor objects that can take on multiple values. Because the computation graph is formed before execution, models in Keras can be saved, loaded, and visualized. | Keras Subclassing API provides an imperative interface where users must develop everything from scratch. |
| PyTorch [19, 20, 22, 23] | PyTorch is a deep learning library for Python programs developed by Facebook's AI research group. It is known for simplicity, ease of use, flexibility, and dynamic computational graphs. | N/A | PyTorch provides an imperative, Pythonic programming style that makes the framework easy to use. It uses dynamic computation graphs that allow users to change how networks behave on the fly while neural network graphs are formed during runtime. Execution is performed imperatively (concretely), meaning dynamic tensor computations are executed immediately. |
| MXNet [27, 29] | Apache MXNet is an open source deep learning framework that offers a hybrid interface, Gluon, that combines imperative and symbolic modes. MXNet has many language bindings including Scala, Java, and C++. | With Gluon, hybridization of neural networks built imperatively allows them to be symbolically executed with speed optimizations. Gluon allows models to be deployed with symbolic graphs in C++ and Scala. | Gluon allows users to develop and train models imperatively, making it suitable for research and prototyping. Gluon also supports dynamic neural network models that can be built on the go with Python's native control flow. |

Table 1: Overview of surveyed models

### 3.1 TensorFlow

TensorFlow [11, 17, 18, 25, 26] is an open source platform that is able to manage all aspects of machine learning powered applications and is flexible with many low-level and high-level APIs. The Google Brain team released the first version of TensorFlow (TensorFlow 1.x) in 2015 and then an updated version, TensorFlow2.0 in 2019 that aims to ease the process of building machine learning frameworks and is tightly integrated with the Keras API [28]. TensorFlow is a symbolic math library and is well suited for dataflow programming across a variety of tasks. By default, TensorFlow 1.x takes a symbolic approach whereas TensorFlow 2.0 takes an imperative one.

In TensorFlow 1.x, users first construct a symbolic graph and then execute it by creating a session. For example, adding two variables using `tf.add` would create a reference to the resulting tensor.

```
1   x = tf.constant(2, name="x")
2   y = tf.constant(3, name="y")
3   sum = tf.add(x, y, name="sum")
4   print(sum)
        #output: Tensor("sum:0", shape=(), dtype=int32)
```

Figure 5: Summing two numbers in TensorFlow

The code in Figure 5 stores the result of `tf.add(x, y)` in the variable `sum` and then tries to print it. However, at this point, only a simple computational graph has been created, so no data has been injected and no computations have been performed. TensorFlow does not perform the addition of `x` and `y` and store it in `sum`, but rather creates a TensorFlow object for `sum` that knows its value can be calculated by summing `x` and `y`. As a result, when line 4 tries to print this symbolic variable, it prints the string representation, `Tensor("sum:0" shape=() dtype=int32)`. In order to print the concrete value of `sum`, a session must be created and run.

```
1   x = tf.constant(2, name="x")
2   y = tf.constant(3, name="y")
3   sum = tf.add(x, y, name="sum")
4   sess = tf.Session()
5   result=sess.run(sum)
6   print(result)
        #output: 5
7   sess.close()
```

Figure 6: Summing two numbers in TensorFlow with a session creation

The revised code in Figure 6 will now print 5, rather than the tensor string representation. TensorFlow 1.x supports only static computation graphs, which can be viewed using the Tensorboard tool [29]. Graphs in TensorFlow 1.x are built and then executed later by creating a session. Sessions only evaluate the tensor they're running on and the nodes it depends on. For example, let us add a multiplication with a new variable to the above example and try to get its result with the session built for `sum`:

```
1   x = tf.constant(2, name="x")
2   y = tf.constant(3, name="y")
3   z = tf.constant(4 ,name="z")
4   sum = tf.add(x, y, name="sum")
5   product = tf.multiply(y, z, name="product")
6   sess=tf.Session()
7   result=sess.run(sum)
8   print(result)
        #output: 5
9   print(product)
        #output: Tensor("product:0", shape=(), dtype=int32)
10  sess.close()
```

Figure 7: Performing two mathematica operations in TensorFlow but only creating a session for one

In Figure 7, although line 8 would print the number 5, line 9 would print the string representation for the tensor, `product`. Since no session is built for `z`, the multiplication is not performed and only `sum` and the nodes it depends on (`x` and `y`) are evaluated. Symbolic execution in TensorFlow means that models are built as data structures such as graphs.

While TensorFlow 1.x separates the building and execution of neural networks with sessions, TensorFlow 2.0 consolidates the two for a much more efficient execution. With the upgrade to TensorFlow2.0 [28], operations are executed imperatively by default, so users no longer have to manually enable eager execution (TensorFlow Eager) [30, 31, 32] which provides an imperative interface to TensorFlow. TensorFlow 2.0's imperative execution means that it does not build graphs and behaves like NumPy or PyTorch. We can compare Figure 5 to its equivalent TensorFlow 2.0 version:

```
1   x = tf.constant(2, name="x")
2   y = tf.constant(3, name="y")
3   sum = tf.add(x, y, name="sum")
4   print(sum)
        #output: tf.Tensor(5, shape=(), dtype=int32)
```

Figure 8: Summing two numbers in TensorFlow 2.0

Due to the default imperative (eager) execution of TensorFlow 2.0, the numerical value of `sum` can be printed without having to create a session, as the operations are executed immediately. Eager execution is able to support dynamic models with Python control flow and is notable for doing well in terms of prototyping and fast debugging due to intermediate run-time errors and Python tools [32]. Eager execution was introduced for TensorFlow to improve its ease of use for new developers and for research and development.

TensorFlow2.0 supports Keras Subclassing, an imperative API, as well as Keras Sequential and Keras Functional, two symbolic APIs, allowing developers to choose which style best suits their needs. Furthermore, the two API styles are interoperable such that models of one type can be nested in or used as a layer in a model of the other type.

## 3.2 Keras

Keras [21, 24, 25, 26] is a high-level API that requires a backend framework to run. It typically runs on top of TensorFlow 2.0, functioning as a wrapper to TensorFlow's framework. Keras can be used to define machine learning models, train them, and make new predictions with them. While TensorFlow is more geared towards machine learning applications, Keras is designed for deep neural networks [25]. In addition, TensorFlow supports both high and low level APIs whereas Keras only supports high-level APIs, meaning it does not have to deal with low-level computations. Although TensorFlow and Keras are tightly integrated, Keras users are not limited to TensorFlow as the Keras API is able to work across several ML frameworks and libraries.

Keras is advantageous for being easily modularized, user friendly, and flexible. Keras is built in Python and has a plug-and-play framework that makes it easy for people to get started. The framework is also able to provide users with clear and effective feedback for errors, and users do not have to code Tensors themselves as they would in TensorFlow because Keras works with Pandas datasets. In addition, Keras allows users to write custom blocks and create new layers, loss functions, metrics, and models. Keras models are created using the Sequential model, the Functional API, or Model subclassing.

Keras Sequential [33] and Keras Functional [34] are two symbolic model-building APIs. Keras Sequential is suited for layers organized as stacks and Keras Functional for those organized as directed acyclic graphs (DAGs). Symbolic models defined in the Sequential and Functional APIs have lower conceptual overhead and are good for copying and cloning. They're also notable for feeling similar to imperative styles of development, making it easy to use for those who are more familiar with imperative APIs.

With symbolic programs, models are translated into graphs behind the scenes and are represented using symbols which later become concrete. Keras implements tensors which are TensorFlow symbolic objects that don't hold defined values [35]. Since each tensor can take on multiple possible values, networks in Keras read input variables and then can form predictions. This symbolically defined model used by Keras makes it possible to run neural networks without having to give values for the input variables in the first instance. When the computation graph is obtained before execution, models can be saved, loaded, and visualized.

```
model.save('path/to/location')
model = model.load_model('path/to/location')
tf.keras.utils.plot_model(model, to_file=img_file, show_shapes=True)
```

Figure 9: Saving, loading, and visualizing a Keras model

Keras Subclassing [10] is an imperative API and is one of the ways to develop models imperatively in TensorFlow 2.0. As an imperative API, it approaches the model from an object-oriented perspective. With Subclassing, users must implement everything from scratch, allowing for more customization and making it more hackable and thus preferable for those working with complex or research purposes.

**3.3 PyTorch**

Pytorch [19, 20, 22, 23] is an open source Python deep-learning library developed by Facebook's AI research group that can be used to develop and train neural networks. While Tensorflow is more commonly used for production, Pytorch is mainly used for research purposes largely due to its ease of use. PyTorch provides a Pythonic programming style and dynamic tensor computations are immediately executed with automatic differentiation and GPU acceleration. At the same time, it aims to provide fast performance which is achieved by being written mostly in C++.

PyTorch's imperative programming style does not use symbolic graphs but rather supports dynamic computation graphs [44]. Unlike TensorFlow's static graphs, dynamic computation graphs allow users to generate graphs and change how networks behave on the fly while neural network graphs are formed during runtime. By allowing computational graphs to be changed during runtime, Pytorch is able to handle situations where the memory requirement is unknown. With dynamic computation graphs, a new computational graph is created at each forward pass.

While symbolic execution uses references, PyTorch's imperative methods mean that operations directly modify data and variables hold actual values. For example, to perform the addition operation that was done in Figures 5 and 8 in PyTorch would look as follows:

```
1   x = torch.Tensor(2)
2   y = torch.Tensor(3)
3   sum = x+y
4   print(sum)
        # output: [torch.FloatTensor of size 1]
```
Figure 10: Summing two numbers in PyTorch

With the PyTorch code in Figure 10, the addition of `x` and `y` has already been performed, as computations are completed line by line as they appear in code, similar to Python program execution. To print out the value of `sum` in Figure 10, users can convert the tensor to NumPy using `sum.numpy()` or use `sum.item()` to retrieve the Python number. These examples also highlight the different API and programming styles of PyTorch and TensorFlow. Compared to TensorFlow, PyTorch tends to feel more intuitive, especially for those who are familiar with imperative programming languages such as Python. For example, mathematical operations in Pytorch can be written using their respective signs, `+` or `*`, whereas in TensorFlow, users would have to use the `add()` or `multiply()` functions. PyTorch aims to provide usability and speed by following the style of the imperative programming paradigm while also supporting hardware accelerators.

**3.4 MXNet**

Apache MXNet [27, 29] is an open-source deep learning framework that trains and deploys deep neural networks. Its scalability allows for fast model training, and it supports a flexible programming model. It has a multi-language library including C++, Python, Julia, Clojure, JavaScript, R, and Scala and is able to support both imperative and symbolic programming interfaces.

MXNet is notable for its flexible Gluon API [6, 7] which combines symbolic expression with tensor computation, enabling a hybrid programming interface that allows users to build and develop models with imperative programming and then convert them to symbolic graphs for performance optimizations. It works by having the developer first build and prototype in the imperative mode using a programming language of their choice. After the model is done, hybridization is performed to switch to the symbolic mode and run the model as a symbolized graph, thus making it easier to be exported and loaded elsewhere for modifications. Gluon also allows users to define dynamic neural networks that can be built using any structure and any of Python's native control flow.

In MXNet, models constructed using the HybridSequential and HybridBlock classes can convert programs from imperative to symbolic [38]. Networks in Gluon are built using Blocks, and a Block that can be converted into a symbolic graph is what characterizes HybridBlocks [37]. HybridBlocks are capable of running not only imperatively, but also symbolically. In the imperative mode, functions and computations act on real inputs whereas the symbolic mode works with placeholders. A sequence of HybridBlocks makes up a HybridSequence. The following code [36] constructs a network using HybridSequential:

```
import mxnet as mx
from mxnet.gluon import nn
from mxnet import nd

def get_net():
    net = nn.HybridSequential()
    with net.name_scope():
        net.add(nn.Dense(256, activation="relu"))
        net.add(nn.Dense(128, activation="relu"))
        net.add(nn.Dense(2))
    net.collect_params().initialize()
    return net

# forward
x = nd.random_normal(shape=(1, 512))
net = get_net()
print('=== net(x) ==={}'.format(net(x)))

    # output:
    # === net(x) ===
    # [[ 0.08827585  0.0050519 ]]
    # <NDArray 1x2 @cpu(0)>

# hybridize
net.hybridize()
print('=== net(x) ==={}'.format(net(x)))

    # output:
    # === net(x) ===
```

```
        # [[ 0.16526183 -0.14005636]]
        # <NDArray 1x2 @cpu(0)>
```
Figure 11: Creating and hybridizing a network in MXNet

Figure 11 exemplifies the hybridization of a network. Calling hybridize() activates HybridBlock which creates a symbolic graph that represents the forward computation and caches it. Hybridization's ability to improve performance by treating the model as a symbolic program rather than an imperative one is demonstrated in Figure 12 [36].

```
from time import time
def bench(net, x):
    mx.nd.waitall()
    start = time()
    for i in range(1000):
        y = net(x)
    mx.nd.waitall()
    return time() - start

net = get_net()
print('Before hybridizing: %.4f sec'%(bench(net, x)))
net.hybridize()
print('After hybridizing: %.4f sec'%(bench(net, x)))

# Before hybridizing: 0.4344 sec
# After hybridizing: 0.2230 sec
```
Figure 12: Measuring the speed before and after hybridization

Hybridization [36] provides a significant speedup in performance. Performing 1000 forward propagations through the network after hybridization cuts the speed in half. Gluon's ability to convert programs into symbolic programming allows it to meet the requirements of product-level computing performance and development. For example, users can train a complex model imperatively with Python, but then deploy with a symbolic graph in C++. Thus, the Gluon package makes prototyping, building, and training easier without compromising on speed.

This process eases development and debugging for users who are more familiar with imperative programming but do not want to compromise on speed as models are hybridized, compiled, and optimized with the backend symbolic engine. Moreover, because the computation structure can be specified by the user, more optimization opportunities can be discovered as clear boundaries are placed on the graph. The Gluon API is often the preferred way to use MXNet as it aims to provide the flexibility and ease of imperative programming when building and debugging the model, but also the superior speed from symbolic programming in converting and optimizing it.

## 4 Comparing Symbolic and Imperative Techniques

Surveying these deep learning frameworks demonstrates the ways in which symbolic techniques are used or forgoed, highlighting areas in which the symbolic paradigm is superior to the imperative, and vice versa.

**4.1 Advantages of Symbolic Techniques**

**Inspection.** Models with symbolic API take on a graph-like data structure, allowing it to be summarized or aggregated. For example, in Keras, models can be plotted as an image using the `plot_model()` function, and `model.summary()` can provide a look into a model's layers, weights, and shapes. Because symbolic programs form graphs before execution, they offer support for functions that provide greater details of the models. With imperative programs, because debugging takes place during model execution rather than definition, there is limited ability to inspect the input or check for layer compatibility, moving the troubleshooting burden from the framework to the developer. For example, with TensorFlow's imperative Keras Subclassing, functions such as `model.save(), model.get_config()`, and `clone_model` do not work for the subclassed models, and `model.summary()` yields a list of layers but no information on them. Furthermore, by using symbolic expressions and path conditions, symbolic execution reveals mathematical characterizations of the internal behavior of neural networks [10].

**Reusability.** The consistency and data schema of symbolic APIs make symbolic models well suited for cloning, sharing, and transfer learning, as existing neurons in intermediate layers can be accessed to create new models. Furthermore, models from symbolic APIs are more easily serialized and thus transportable across languages because they can be serialized from one language and deserialized and executed in another as long as the languages have the same symbolic programming syntax. Because imperative models do not have the consistent API that is provided by symbolic ones, they are more difficult to reuse as intermediate layers and neurons are inaccessible.

**Debugging.** Developers can splice layers together and run layer compatibility checks, similar to type-checking in the compiler. This helps reduce errors since troubleshooting is performed during model definition and customization, rather than execution. As a result, it ensures that compiled models will work, making troubleshooting easier and iterations faster. Symbolic execution also helps minimize errors because it yields an effective bug finding technique; a solver can be used to find concrete values that lead to errors based on the symbolic inputs the programs are evaluated on.

**Optimizations.** By not forming a dataflow graph, frameworks with imperative techniques cannot apply optimizations that are available to those with a symbolic graph execution model. One of these optimizations is operation folding. In the following MXNet example [16], the multiplication in line 4 and addition in line 5 can be combined into one operation, b+a+1.

```
1 import numpy as np
2   a = np.ones(10)
3   b = np.ones(10) * 2
4   c = b * a
5   d = c + 1
```

Figure 13: Mathematical computation in MXNet

This operation folding can be illustrated as follows [16]:

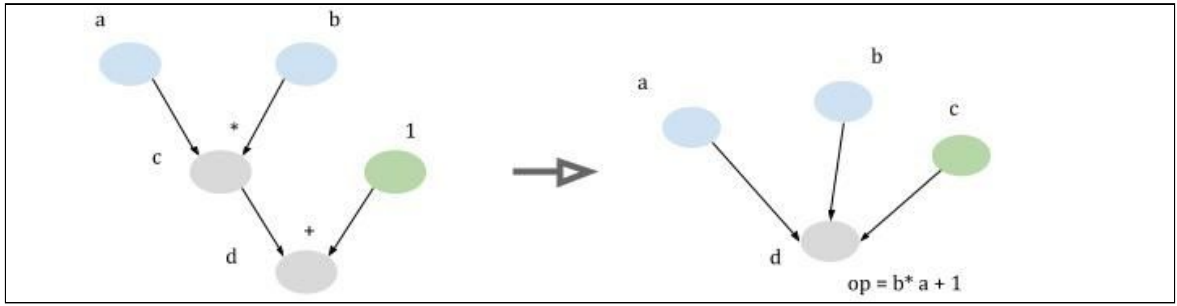

Figure 14: Graph illustration of operation folding based on Figure 13

Because a compiler can view the entire code before turning it into machine instructions, it can optimize and rewrite the code. As a result, if the computation runs on a GPU, the number of GPU kernels executed will be cut in half from two to one, improving computation efficiency. Operation folding cannot be done in imperative programs because, unlike symbolic programs, imperative ones do not know which values will be needed in the future. As a result, they keep the values of all the variables in case they are referenced later on. This leads to another optimization limitation of imperative programming which is that reducing memory requires line by line inspection. Figure 15 demonstrates a scenario where symbolic programs would emerge as more memory and speed efficient [16].

|  | Imperative | Symbolic |
|---|---|---|
| ```
1  import numpy as np
2    a = np.ones(10)
3    b = np.ones(10) * 2
4    c = b * a
5    d = c + 1
``` | 4  (# of arrays)<br>10 (size of array)<br>x  8  (bytes per array cell)<br>-------------<br>320 bytes | 2  (# of arrays)<br>10 (size of array)<br>x  8  (bytes per array cell)<br>-------------<br>160 bytes |

Figure 15: Memory usage in imperative and symbolic modes

In Figure 15, assuming each array cell requires 8 bytes of memory, an imperative program would need to allocate 320 bytes of memory whereas a symbolic program would only need 160. Imperative programs don't know whether variables will be used later on so it keeps all of them and must allocate memory at each line. As a result, an imperative approach would allocate 4 arrays of size 10, totalling to 4*10*8=320 bytes. On the other hand, creating a computation graph would be able to determine areas of memory that can be released or recycled. In Figure 15, a symbolic program, when compiled, would notify the system that only the value of d is needed, so memory initially allocated to other variables can be reused. Bits for variable b may be recycled to store c, and then bits for c recycled for d, ultimately ending with a memory usage of 2*10*8=160 bytes.

With a symbolic API, memory for variables can be reused or never allocated since the computational graph is recorded before it is bound to inputs and executed. However, symbolic



programs' ability to reuse memory for in-place computation becomes problematic if users need to access intermediate variables, such as c in Figure 15. In such cases, imperative would be superior as intermediate variables can be accessed by running code in a Python console.

## 4.2 Disadvantages of Symbolic Techniques

**Flexibility.** Symbolic APIs do not work well for tree shape, dynamic, and recurrent neural networks since compilation forces the model to become static. Moreover, user experience can be complicated by the separation of building and executing a symbolic graph, as defining neural networks does not run numerical computations and constructed graphs are later executed by separate functions.

**Usability.** The imperative style is superior to symbolic in regards to its more natural feel. Users often find imperative programming more intuitive and easier to write in, and it is often compared to writing in Python. Symbolic programs do not always have access to familiar Python constructs such as iteration, making the imperative style potentially better for someone new to deep learning or someone who wants to be more experiential and work with new ideas.

The following MXNet code examples illustrate how someone unfamiliar with deep learning may feel more comfortable dealing with an imperative program over a symbolic one [38].

| Imperative programming (NDArray API) | Symbolic programming (Symbol API) |
|---|---|
| ```
1 import mxnet.ndarray as nd
2 a = nd.ones((4, 4))
3 b = nd.ones((4, 4))
4 c = a + b
5 print (c.asnumpy ())
``` | ```
1 import mxnet.sym as sym
2 import numpy as np
3 a = sym.Variable ('a', shape=(4, 4))
4 b = sym.Variable ('b', shape=(4, 4))
5 c = a + b
6 exe = c.simple_bind(ctx=mx.cpu())
7 exe.forward(a=np.ones((4, 4)))
8 print(exe.outputs[0].asnumpy())
``` |

Figure 16: Side by side comparison of an imperative and symbolic program in MXNet

It is clear that the imperative version is more concise and readable, especially to developers with minimal experience. Moreover, with the symbolic version, users have the additional work of having to compile the executor (line 6) and then run it (line 7).

**Debugging.** Because imperative programs keep all intermediate values, users are able to print during execution, giving it an edge over symbolic programs when it comes to debugging. In addition, some users prefer the debugging experience in an imperative API over symbolic because symbolic models are compiled as a black box which internally may differ from the syntax used to construct the model. For example, Keras wraps a lot of computational blocks in abstractions, making it difficult to pinpoint the line where the source of an error lies.

## 4.3 Summary of the advantages and disadvantages

| | Symbolic | Imperative | Reasoning |
|---|---|---|---|
| Inspection | ✓ | | Models can be plotted and summarized due to their graph-like structure, |
| | | | and layer information and compatibility can be checked before execution. |
| Reusability | ✓ | | The ability to serialize models and access existing neurons in intermediate layers makes symbolic models easier to clone and transport. |
| Debugging | ✓ | ✓ | With symbolic execution, troubleshooting is performed during model definition rather than execution, ensuring compiled models will always work. Symbolic programming can also make use of a solver to find concrete values that lead to errors based on the symbolic inputs the programs are evaluated on. On the other hand, imperative programming allows for printing during execution, which can aid in the debugging process. |
| Optimizations | ✓ | | Symbolic programs build computation graphs before execution, allowing them to determine which variables will need to be kept and areas where memory can be reused. As a result, they are able to perform optimizations such as operation folding which increases computation and memory efficiency. |
| Flexibility | | ✓ | Imperative APIs are able to handle tree shape, dynamic, and recurrent neural networks, which symbolic APIs cannot due to their static nature. |
| Usability | | ✓ | Imperative programming is pervasive, so many users are already familiar with its style. Due to its relative ease of use and relative readability, it is more appealing to new and experiential users. |

Table 2: Overview of advantages and disadvantages of symbolic and imperative paradigms
* ✓ indicates the framework that exhibits advantages over the other

Symbolic programming tends to stand out for its superiority in inspecting, and reusability, and optimization models, but overall struggles in comparison to the imperative style when it comes to flexibility and usability. Both paradigms have their own advantages when it comes to debugging.

Given the advantages and disadvantages of symbolic and imperative APIs, some deep learning models have attempted to take the best of both paradigms. For example, TensorFlow allows developers to choose between imperative and symbolic APIs, and MXNet takes the simplicity and flexibility of a framework like PyTorch while also allowing network hybridization to take advantage of performance optimizations offered by symbolic graph execution models. As a result, MXNet prides itself on having imperative attributes which make it well suited for experimentation with new or complex neural network models. Similar to MXNet, TensorFlow's recent version also implements computational graphs but uses imperative programming to specify the computation within the graph.

## 5 Challenges and Opportunities

It is a difficult task to fully take advantage of the best of both symbolic and imperative techniques while also limiting their drawbacks. It seems that the challenge lies in obtaining both high performance and good usability and that one must often be sacrificed to an extent in order to achieve the other. However, these challenges have also led to new developments in deep learning, particularly in regards to how symbolic and imperative techniques can be utilized together.

**Switch between paradigms.** Users are often having to learn multiple paradigms or frameworks and switch between them based on their needs. For example, TensorFlow 1.x is symbolic by default, but users can toggle between imperative and symbolic modes. In the later release of TensorFlow 2.0, the default was switched to the imperative but the framework still allows for symbolic programs. Although TensorFlow added support for imperative programming, it kept its symbolic functionalities. The changes to TensorFlow from 1.x to 2.0 demonstrate the appeal of writing in imperative code yet the hesitation to only support one or the other as they each offer different benefits and are preferred for different uses.

**Language support.** Another issue is determining how to harness the usability of imperative programming and strike a medium in which it is user friendly but doesn't support so many languages that it becomes confusing to use. For example, while Gluon's imperative methods are intended to make it easy to learn and use, its support of a wide range of programming languages in the frontend has come to the criticism of users for making it difficult to search for code. The numerous different languages possible for implementation means that although a user's desired implementation may exist, it may not be in a language they are familiar with, thus preventing them from finding and using it.

**Symbolic representation of imperative programs.** Attempts have been made to capture the dynamic characteristics of imperative Python programs in a symbolic dataflow graph. One of these attempts is JANUS [15, 39, 40], a framework that constructs symbolic graphs of imperative deep learning programs. JANUS works by assuming speculative program context from the input program and uses imperative execution to correct any wrong assumptions that could invalidate the symbolic graph. The work on JANUS highlights opportunities for integration of imperative programs and symbolic graphs.

**Concolic testing.** Another potential opportunity for symbolic techniques in deep learning is the combination of symbolic analysis and concrete execution. This technique, called concolic testing [41], is typically used to explore execution paths of software programs and has been considered for use in deep neural networks (DNNs). DeepConcolic [42] proposes a concolic testing approach to DNNs that involves alternating between concrete execution and symbolic analysis. DeepConcolic works by first executing the program with a concrete input, then selecting and symbolically encoding another execution path, and finally using a constraint solver to solve the path's resulting formula to produce a new concrete input. This process of performing concrete runs with symbolic analysis is repeated in an alternating fashion until an acceptable level of structural coverage is obtained.

Concolic testing has the potential to be a good candidate for use with DNNs because DNNs tend to have a high dimensional input space that makes random testing unsuitable. Moreover, symbolic execution isn't capable of covering all the execution paths generated by the commonly used ReLU activation function. However, concolic testing can direct the symbolic analysis to particular execution paths through concrete evaluation of the DNN's properties, thus making it possible to deal with a greater number of execution paths than symbolic execution alone. When dealing with an unsatisfied test requirement x, concrete execution finds a test input within the available test suite that is close to satisfying x, and then symbolic analysis is put into action to find a new test input that does satisfy x. This new test input is then added to the test suite and this process is repeated, gradually increasing the number of test inputs and improving coverage.

**Symbolic execution on imperative programs.** One challenge with symbolic representation is that performance degrades as symbolic graphs grow larger; networks with thousands of neurons are too difficult for current constraint solvers that try to determine the feasibility of symbolically encoded paths. This scalability issue, in addition to neural networks' minimal branching and often non-linearity and the under development of constraint solvers, are a few of the problems with symbolic execution which has led to initiatives like DeepCheck[9]. DeepCheck is an approach to transform neural networks into imperative programs that can be analyzed and validated using symbolic execution. It does this by translating the neural network into an imperative program, performing symbolic execution to obtain path constraints for certain concrete inputs, determining key characteristics from the inputs for classification, and synthesizing adversaries that trick the network into wrong classifications.

## 6 Conclusion

In this paper, we explored how different deep learning frameworks use symbolic techniques, revealing not only the advantages to a symbolic approach, but also the challenges and opportunities that come with them. By looking at TensorFlow and Keras, we learn how deep learning models with symbolic APIs utilize symbolic execution to create models that work with symbolic objects and, as a result, are superior when it comes to inspection, reusing, and optimizing. In contrast, TensorFlow 2.0 and PyTorch exhibit an imperative API with an imperative programming style that is often more flexible and easier to use and experiment with.

It is debated which areas symbolic techniques do better in compared to imperative, and vice versa, adding to the challenge of trying to integrate the two together. MXNet's Gluon API provides a framework that utilizes both symbolic and nonsymbolic techniques in hopes of combining imperative programming's ease of use with symbolic programming's optimization powers. Similar to Gluon, recent approaches such as JANUS, DeepConcolic, and DeepCheck have attempted to take advantage of both paradigms; JANUS creates symbolic graphs of imperative programs, DeepConcolic alternates between using symbolic analysis and concrete execution, and DeepCheck uses symbolic execution to analyze imperative programs. These projects are opening up new ways to utilize symbolic techniques in deep learning.

The increasing demand for deep learning indicates this paper would benefit from surveyance of other popular frameworks such as Caffe [49], ONNX [50], and Theano [51] and the ways in which they choose to use or forgo symbolic techniques. At the same time, the continuous evolution of deep learning calls for continued monitoring of the aforementioned frameworks as well as new emerging frameworks. Continuing development in these frameworks could provide insight into the trends and trajectories of machine learning and reveal weaknesses in current frameworks that are being revised or improved upon by new ones. For the purposes of this paper, it is important to keep the focus on symbolic techniques, particularly whether there are new techniques emerging or being changed and adapted to new uses. By looking at the ways in which deep learning frameworks choose and modify certain symbolic techniques, we can better understand their advantages and disadvantages, ultimately paving the way for more effective uses and adaptations.

**Acknowledgements**

This research received financial support in part from National Science Foundation, Grant #: 1908571 and its Research Experiences for Undergraduates (REU) supplements.